# Steganography in Handling Oversized IP Packets

Wojciech Mazurczyk and Krzysztof Szczypiorski
Warsaw University of Technology, Institute of Telecommunications
Warsaw, Poland, 00-665, Nowowiejska 15/19

**Abstract.** This paper identifies new class of network steganography methods that utilize mechanisms to handle oversized packets in IP networks: IP fragmentation, PMTUD (Path MTU Discovery) and PLPMTUD (Packetization Layer Path MTU Discovery). In particular, we propose two new steganographic methods and two extensions of existing ones. We show how IP fragmentation simplifies utilizing steganographic methods which requires transmitter-receiver synchronization. We present how mentioned mechanisms can be used to enable hidden communication for both versions of IP protocol: 4 and 6. Also the detection of the proposed methods is enclosed in this paper.

Key words: network steganography, IP fragmentation, PMTUD, PLPMTUD

## 1.  Introduction

   Communication network steganography is a method of hiding secret data in users' normal data transmissions, ideally, so it cannot be detected by third parties. Many new methods have been proposed and analyzed, e.g. [1], [3] or [4]. Network steganography methods may be seen as a threat to network security as they may be used as a tool to cause for example confidential information leakage. That is why it is important to identify potential possibilities for covert communication, because knowledge of the information hiding procedure can be used to develop countermeasures.
   Both versions of IP protocol 4 [5] and 6 [9] were designed to be used on various transmission links. The maximum length of an IP packet is 64 kB but on most transmission links maximum packet length is smaller - this limited value characteristic for the specific link is called a MTU (Maximum Transmission Unit). MTU depends on the type of the transmission link e.g. for Ethernet - 1500, wireless IEEE 802.11 - 2300 and PPP (Point to Point Protocol) - 296 bytes.
   There are two possibilities to transmit packets through an end-to-end path that consists of links with different MTUs:
- **Permit to divide oversized packet** to smaller ones. To achieve this mechanism called IP fragmentation [5] has been standardized.
- **Do not allow packet fragmentation** and adjust IP packet size to so called PMTU (Path MTU) – the smallest, acceptable MTU along the entire end-to-end path. For this purpose two methods have been proposed PMTUD (Path MTU Discovery) [6] for IPv4 and [7] for IPv6 and PLPMTUD (Packetization Layer Path MTU Discovery) [8], which is enhancement of previous method for both versions of IP protocol.

   Mechanisms for handling oversized packets like IP fragmentation, PMTUD or PLPMTUD are needed and used in network scenarios where in the end-to-end path intermediate links have smaller MTUs than the MTU of the end links. Typical network scenarios that require dealing with oversized packets:
- Using various tunneling protocols like GRE (Generic Routing Encapsulation), IPSec (IP Security), and L2TP (Layer Two Tunneling Protocol) which add headers and trailers which result in reduced effective MTU.
- Using PPPoE (Point to Point Protocol over Ethernet) with ADSL (Asymmetric Digital Subscriber Line). PPPoE has 8 bytes header thus it reduces the effective MTU of the Ethernet to 1492.
- Using MPLS over Ethernet.
- Connections between endpoints in Token Ring or FDDI networks, with greater MTU, which have an Ethernet link between them, with lower MTU, and other similar cases.

   The objectives of this paper are to:
- Describe mechanisms used to handle oversized packets in IPv4 and IPv6 networks.
- Present exiting network steganography methods that utilize these mechanisms.

- Propose two new steganographic methods and two extensions of existing ones All presented steganographic methods may be applied to both versions of IP protocol (4 and 6).

The rest of the paper is as follows. Section 2 describes existing mechanisms for handling oversized packets for IPv4 and IPv6 protocols. In Section 3 network steganography methods that utilize IP fragmentation mechanism are presented. Section 4 includes detailed description of new proposed information hiding methods and their potential detection. Section 5 concludes our work.

## 2. Overview of Mechanism for Handling Oversized IP Packets

### 2.1 IP Fragmentation

To accommodate MTU differences on links in end-to-end path in IP fragmentation, intermediate nodes are allowed to fragment oversized packets to smaller ones. Then receiver or some other network node (e.g. router) is responsible for reassembling the fragments back into the original IP packet.

IP fragmentation mechanism involves using the following fields of the IPv4 header: *Identification*, *Fragment Offset* fields, along with the MF (More Fragments) and DF (Don't fragment) flags (Fig. 1). It also needs to adjust values in *Total Length* and *Header Checksum* fields for each fragment to represent correct values. The above header fields are used as follows:
- **Identification** (16 bits) is a value assigned by the sender to each IP packet to enable correct reassembling of the fragments (each fragment has the same *Identification* value). The value used in IP *Identification* header field must uniquely identify an IP packet for a certain amount of time [5].
- **Fragment Offset** (13 bits) indicates which part of the original packet fragment carries. Value in this field should be multiple by eight to calculate real amount of bytes of original packet.
- **Flags field** (3 bits) contains control flags. Bit '0' is reserved, can not be used and is always set to 0. Bit '1' is the DF bit. If its value is set to 0 fragmentation can take place if necessary. If it is set to 1 then fragmentation of the packet is impossible. Bit '2' is the MF – if it is set to 0 and *Fragment Offset* is different from 0, then this means last fragment and if it is set to 1 it indicates that more fragments are expected to be received.

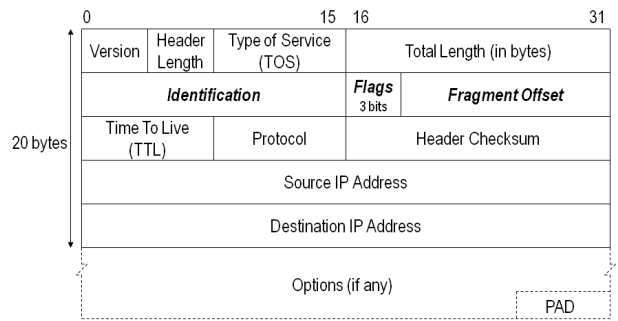

**Fig. 1** The IPv4 protocol header (bolded are fields used by IP fragmentation)

Similar mechanism is used in version 6 of IP protocol, where *Fragment* extension header is used to perform fragmentation (Fig. 2). What differs IPv6 from IPv4 fragmentation is that it is always performed by the sender and reassembly process have to take place only in the receiver and not in some intermediate node.

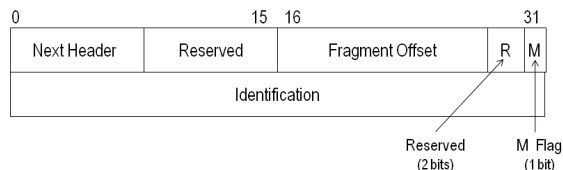

**Fig. 2** IPv6 *Fragment* header extension

The example of the IP packet fragmentation for IPv4 is presented in Fig. 3. Original packet which size is 5140 bytes is divided into four fragments of maximum 1500 bytes.

**Original IP Packet**

| Sequence | Identifier | Total Length | DF | MF | Fragment Offset |
|---|---|---|---|---|---|
| 0 | 345 | 5140 | 0 | 0 | 0 |

**IP Fragments**

| Sequence | Identifier | Total Length | DF Flag | MF Flag | Fragment Offset |
|---|---|---|---|---|---|
| 0-0 | 345 | 1500 | 0 | 1 | 0 |
| 0-1 | 345 | 1500 | 0 | 1 | 185 |
| 0-2 | 345 | 1500 | 0 | 1 | 370 |
| 0-3 | 345 | 700 | 0 | 0 | 555 |

**Fig. 3** Example of IP fragmentation

There are several issues that make IP Fragmentation in IPv4 networks undesirable because it lowers the efficiency and reliability of communication. Fragmentation causes serious overhead for the receiver because while reassembling the fragments the receiver must allocate memory for the arriving fragments and after all of the fragments are received they are put back into original IP packet. While it is not an issue for a host as it has the time and memory resources to devote to this task, reassembly may be very inefficient on intermediate nodes (e.g. routers). Router is not able to determine the size of the original IP packet until the last fragment is received, so while reassembling it must assign a large receiving buffer.

Another fragmentation issue involves handling dropped fragments. If one fragment of an IP packet is dropped, then the entire original IP packet must be resent (all fragments).

Firewalls and NATs (Network Address Translation) may have trouble processing fragments correctly and in effect drop them. If the IP fragments are out of order, a firewall may block the non-initial fragments because they do not carry the information that would match the packet filter. This would mean that the original packet could not be reassembled by the receiving host. Similar problem may occur with NAT as it has problems with interpreting the IP fragment if it comes out of order.

**2.2 PMTUD (Path MTU Discovery)**

PMTUD was standardized and published for IPv4 in 1990, but it did not become widely deployed for the next few years – currently PMTUD is implemented in major operating systems (Windows, Unix, Linux) – in 2002 about 80% - 90% of endpoints on the Internet were using this mechanism. As mentioned in the introduction this mechanism was developed to avoid fragmentation in the path between the endpoints. Similar to IPv4 PMTUD mechanism was also developed and standardized for IPv6 [7].

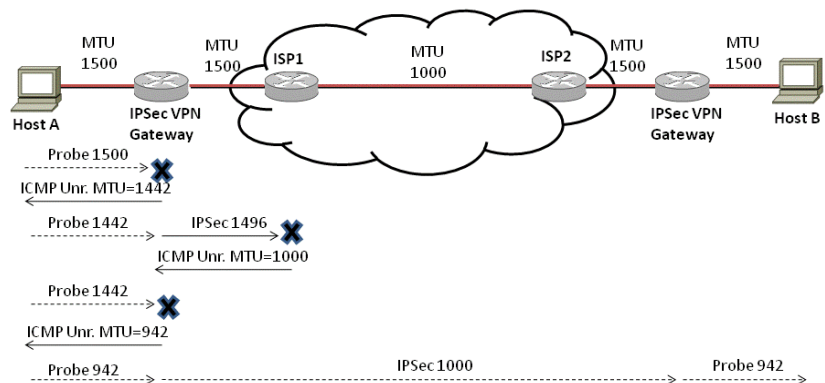

**Fig. 4** PMTUD example

PMTUD is used to dynamically determine the lowest MTU along the end-to-end path between packet's sender and receiver. Instead of fragmenting packet, an endpoint tries to find out the largest possible size of the packet that can be sent to a specific destination. An endpoint finds out the correct packet size associated with a specific path by sending packets with different sizes. Packets used by PTMUD are called probe messages and they have DF (Don't Fragment) flag set in the IP protocol header. Their size is initially set to the senders link MTU. While sender generates probes he/she responds to possible ICMP (Internet Control Message Protocol) error reports that indicate a low MTU is present along the connection path. Sender gets back a notification saying what size will be suitable. The notifications are requested by setting the DF flag in outgoing packets. For IPv4 the notifications arrive as ICMP messages known as "Fragmentation required, and DF flag set" (ICMP type 3, code 4), for IPv6 it is "Packet too big" message from ICMPv6 protocol [10]. PMTUD is working continually during connection because the path between sender and receiver can changed (e.g. because of link failure).

The example of how PMTUD is used to determine correct MTU value along the connection path is illustrated in Fig. 4. Host A sends packet to host B which size is set to 1500 bytes (default Ethernet MTU). The packet will be transmitted with use of IPSec tunnel, which begins at first router. Because the next link MTU is also 1500 bytes and IPSec adds 54 bytes overhead then total packet size exceeds admissible value. Thus the packet is dropped and ICMP message is sent back to the host A with suitable MTU for the next link. Then host A retries sending the packet by reducing its size to 1442 bytes to meet the limit so packet can successfully traverse through first router. However, the link after next router has MTU of 1000 bytes so the packet is dropped and ICMP message is sent in host A direction but it is filtered out by first router. After the timeout expires host A retransmits the packet and receives ICMP message which indicates necessity to decrease packet's size to 942 bytes. This last MTU value is then used to successfully exchange packets with host B.

It must be noted that there are security issues connected with using PMTUD. In particular, sometimes network administrators treat ICMP traffic as dangerous and block it, disabling possibility of using path MTU discovery. Other potential issues for TCP protocol are described in [11].

### 2.3 PLPMTUD (Packetization Layer Path MTU Discovery)

To alleviate issues related with using ICMP traffic for PMTUD, enhancement called PLPMTUD was developed and standardized in [8]. What differs PLPMTUD from PMTUD is that receiving probes messages are validated at the transport layer. It does not rely on ICMP or other messages from the network, instead it learns about correct MTU by starting with packets which size is relatively small and when they get through with progressively larger ones. In particular, PLPMTUD uses a searching technique to find optimal PMTU. Each probe narrows the MTU search range. It may raise the lower limit on a successful probe receipt or lower the upper limit if probe fails. The isolated loss of a probe message is treated as an indication of an MTU limit and transport layer protocol is permitted to retransmit any missing data.

## 3. Related Work

To authors best knowledge, there are no steganographic methods proposed for PMTUD and PLPMTUD mechanisms.

There are few existing methods for IPv4 that utilize IP fragmentation mechanism and fields in IP header related to it. Rowland [1] proposed multiplying each byte of the hidden data by 256 and inserts it directly into *Identification* header field.

Cauich et al. [14] described how to use *Identification* and *Fragment Offset* fields to carry hidden data between intermediate nodes but under condition that the packet is not fragmented. Additionally in selected packet reserved flag is used to mark packet so that the receiver can distinguish between real and covert fragments.

Murdoch et al. [4] proposed transmitting hidden information by modulating the size of the fragments to match the hidden data inserted into *Fragment Offset* field.

Ahsan and Kundur [12] proposed steganographic method that use IP fragmentation fields. It utilizes high eight bits of the IP Identification to transmit covert data and the low eight bits are generated randomly. The same authors in [13] described a method that uses DF flag as a covert data carrier. If the sender knows the correct MTU for the end-to-end path to the receiver and issues packets which size is less than MTU then DF can be set to arbitrary values.

For IPv6 protocol Lucena et al. [15] identified four network steganographic methods based on Fragment header extension. Two methods use reserved fields to carry steganogram and one next header field. Fourth steganographic method is based on fake fragments insertion. In this case all fields of the fragment header may be used to covert communication. To avoid having inserted fragment included in the reassembly process of the original IP packet, authors propose two solutions: first is based on inserting an invalid value in *Identification* field in *Fragment extension* header thus the receiver will not use this fragment for reassembly, second – inserting overlapping *Fragment Offset* value that causes data to be overwritten during reassembly. Fake fragments carry hidden data only in certain header fields.

## 4. Proposed Methods: Communication Scenarios, Functioning and Detection

Every steganographic method should be analyzed in terms of steganographic bandwidth and risk of hidden communication disclosure. Steganographic bandwidth may be expressed by means of RBR (Raw Bit Rate), which is defined as a total number of steganogram bits transmitted during one time unit [bit/s] or equivalently by

PRBR (Packet Raw Bit Rate) which is defined as a total number of steganogram bits transmitted in single packet used during the hidden communication process [bit/packet]. Some steganographic methods are trivial to detect (e.g. those which simply modifies header fields) but for others the steganalysis may be hard to perform. Thus for each proposed steganographic solutions potential detection methods must be analyzed in details.

In general, there are four communication scenarios for network steganographic exchange possible. The first scenario (1) in Fig. 5, is most common: the sender, who is also a Steganogram Sender (SS) and the receiver, who is also a Steganogram Receiver (SR) establish a connection while simultaneously exchanging steganograms. In the next three scenarios (marked 2-4 in Fig. 5) only a part of the connection end-to-end path is used for hidden communication as a result of actions undertaken by intermediate nodes; the sender and/or receiver are, in principle, unaware of the steganographic data exchange.

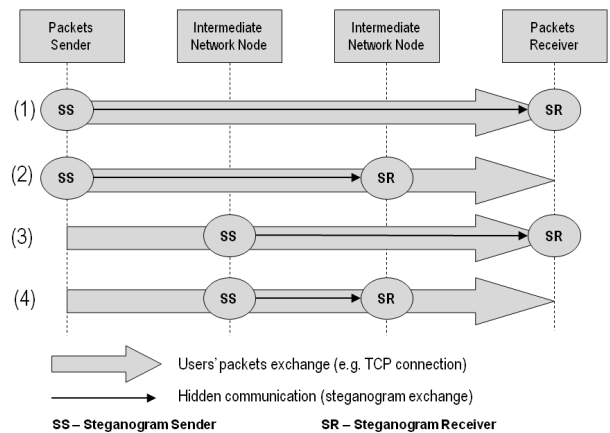

**Fig. 5** Hidden communication scenarios

Hidden communication scenarios presented above differ in steganalysis, in particular, the scenario 4 is harder to detect, because the network node which analyses traffic for hidden communication called warden [20] is usually placed at the edge of source or destination endpoints (sub)network.

**4.1 IP Fragmentation**

For IP fragmentation mechanism we propose new steganographic method (F1) and two enhancement of the previously proposed ones (F2 and F3). Moreover, we also show how IP fragmentation simplifies usage of existing steganographic methods that require transmitter-receiver synchronization (F4-6). Steganographic methods that may be used for IP Fragmentation can be classified as presented in Fig. 6.

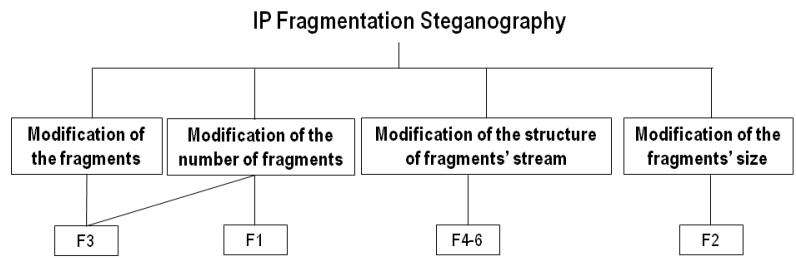

**Fig.6** Classification of IP Fragmentation steganographic methods

Each of presented methods may be utilized for IPv4 and IPv6 protocols for each scenario from Fig. 5. However, for IPv4 fragmentation and fragments reassembly may be performed by intermediate nodes as well as by the sender and/or receiver. This may limit the steganogram exchange only to the fragmenting and assembling nodes. For IPv6 fragmentation may be performed only for sender and assembly by the receiver of the packets so there is no such limitation.

**Steganographic method F1**

In this method SS (Steganogram Sender) must be the source of the fragmentation. SS inserts single bit of hidden data by dividing original IP packet into the predefined number of fragments. For example, if the number of the fragments is even then it means that binary "0" is transmitted and in other case binary "1" (Fig. 7).

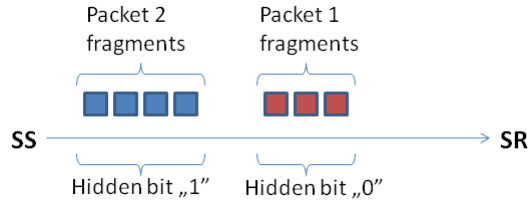

**Fig. 7** F1 steganographic method example

After reception of the fragments SR uses the number of the fragments of each received IP packet to determine what hidden data was sent.

Potential steganographic bandwidth for this method is PRBR = 1 bit/packet.

Detection of this method may be hard to perform. Statistical steganalysis based on number of fragments can be performed to detect irregularities in number of the fragments. The best method to make hidden communication unavailable is to reassembly original IP packet in some intermediate node which is responsible for detecting steganographic communication (warden [20]), refragment it randomly and send to the receiver.

**Steganographic method F2**

The main idea of this method is to divide a packet into fragments and insert hidden information by modulating the values that are inserted into *Fragment Offset* field. As mentioned in Section 3, Murdoch et al. [4] proposed inserting steganogram directly into *Fragment Offset* field and modulate the size of the fragment to match this value. Such approach can cause high irregularities in fragments sizes which may be easily detected. We propose enhancement of this method which has lower steganographic bandwidth but is harder to detect.

F2 method works as follows. SS must be the source of the fragmentation. SS inserts single bit of hidden data by intentionally modulating the size of each fragment of the original packet in order to obtain fixed values in *Fragment Offset* field. For example, even offset means transmitting binary '1', odd offset – binary '0'. "Steganographic" fragmentation of the exemplary IP packet which was introduced in Fig. 3 is presented in Fig. 8.

| Sequence | Identifier | Total Length | DF Flag | MF Flag | Fragment Offset | Hidden data |
|---|---|---|---|---|---|---|
| 0-0 | 345 | 1300 | 0 | 1 | 0 | - |
| 0-1 | 345 | 1340 | 0 | 1 | 160 | 1 |
| 0-2 | 345 | 1340 | 0 | 1 | 325 | 0 |
| 0-3 | 345 | 1220 | 0 | 0 | 490 | 1 |

**Fig. 8** Example of F2 steganographic method

After successful reception of the fragments SR extracts hidden data based on the values from *Fragment Offset* field.

Steganographic bandwidth for this method is $PRBR = N_F - 1$ [bit/packet], where $N_F$ denotes number of fragments of the packet.

Steganalysis in case of F2 is harder than in case of method proposed by Murdoch but hidden communication still can be uncovered, because usually all the fragments except last one have equal sizes (see Fig. 3). Thus, if there are any irregularities in fragments sizes then steganographic communication may be uncovered. However, this method may be further improved, so the detection is more difficult to perform. We may influence the size of the fragments in such a manner that all fragments except last one would have the same length and the value in *Fragment Offset* field in last fragment is modulated to achieve even or odd value. In this case the hidden communication may not be detected at all as this fragmented packet will be similar to other ones.

Steganographic bandwidth for this improved method will be lower than for above method and will be equal PRBR = 1 bit/packet.

Detection of this method may be hard to perform. Statistical steganalysis based on fragments sizes can be performed to detect irregularities. The best method to make the hidden communication unavailable is the same as in case of method F1.

**Steganographic method F3**

Proposed method is enhancement of Lucena et al. [15] work for IPv6 Fragmentation where they proposed to generate fake fragments. As mentioned in section 3 two solutions to distinguish fake fragments from the legitimate were presented – first is based on inserting an invalid value in *Identification* field in *Fragment extension* header, second – inserting overlapping *Fragment Offset* value that causes data to be overwritten during reassembly. Fake fragments carry hidden data only in certain header fields. However, described methods may be easy to uncover because the warden can monitor all the fragments sent and determine potential anomalies like overlapping offsets or single, unrelated fragments. Our proposition is to use legitimate fragment with steganogram inserted into payload for higher steganographic bandwidth and harder detection.

F3 method works as follows. SS must be the source of the fragmentation. SS while dividing the packet, instead of inserting user data into the payload of selected fragment he/she inserts steganogram. The problem with such approach is to properly mark fragments used for hidden communication so the receiver can extract it in a way that will not interfere with reassembly process. We propose the following procedure to make the selected fragment distinguishable from other yet hard to detect. Let us assume that sender and receiver share secret Steg-Key (*SK*). For each fragment chosen for steganographic communication the following hash function (*H*) is used to calculate Identifying Sequence (*IS*):

$$IS = H(SK \parallel Fragment\ Offset \parallel Identification) \qquad (4\text{-}1)$$

where *Fragment Offset* and *Identification* denote values from these IP fragment header fields and $\parallel$ bits concatenation function. For every fragment used for hidden communications the resulting *IS* will have different value due to the changing values in *Fragment Offset*. All *IS* bits or only selected ones are distributed across payload field in predefined manner. Thus the receiver for each fragment based on *SK* and values from the IP header can calculate appropriate *IS* and check if it contains steganogram or user data.

For each incoming fragment SR calculates *IS* and checks if it carries steganogram. If the verification is successful then rest of the payload is considered as hidden data and extracted. Then SR skips this fragment in reassembly process of original IP packet.

Steganographic bandwidth for this method may be expressed as

$$PRBR = N_F \cdot F_S \quad [bits/packet] \qquad (4\text{-}2)$$

where $N_F$ denotes number of fragments and $F_S$ the size of the fragment payload.

Fig. 9 illustrates example of the proposed steganographic method. IP packet with ID 345 is divided into four fragments (F1-F4). Fragment F2 is used for steganographic purposes, so inside its payload steganogram is inserted together with correct *IS*. Values in *Fragment Offset* and *Identification* remain the same as in other legitimate fragments. While reassembling original packet, receiver merges payloads P1, P2 and P3, omits fragment F2 and use it only to extract steganogram.

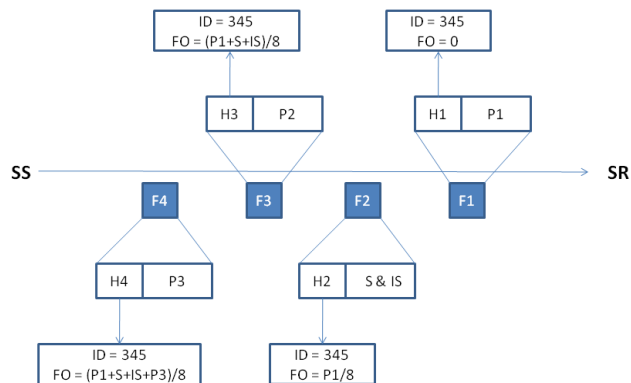

**Fig. 9** F3 steganographic method example (H – header, P – payload)

Method F3 is hard to detect because legitimate fragments are used as hidden data carriers. The best method to make the hidden communication unavailable is the same as in case of methods F1 and F2.

**Steganographic methods F4-6**

Fragments that are created during fragmentation process may be treated as numbered stream of the packets, because *Identification* and *Fragment Offset* fields uniquely identify each piece and allow their correct placement during reassembly process. That is why, for IP fragmentation mechanism existing network steganographic methods proposed for such numbered data may be utilized. These are: intentional changing sequence of the packets, modifying inter-packet delays and introducing intentional losses. What is common to these methods is sender-receiver synchronization requirement. We show that for fragmentation process this requirement is not longer valid, so the deployment of these methods is easier – synchronization is not needed because one packet fragmentation may be treated as one synchronization period. The lack of requirement for sender-receiver synchronization makes these methods easier to implement.

Intentional changing sequence of the packets for transmitting covert data was proposed in [16, 17]. These methods may be applied to IP Fragmentation, especially if the number of fragments is high by sending fragments in a predefined fashion. In Fig. 3 four fragments were created and *Fragment Offset* values decide of their sequence. So sending fragments in the sequence 0, 1, 2, 3 may be interpreted as binary '1' and the reverse order as binary '0'.

Network steganography method that modifies inter-packet delay was presented in [18]. Such approach may be successfully utilized for IP fragmentation and for example work as follows. During fragmentation of one IP packet, fragments are generated at one rate (it may mean sending hidden binary '1') and while dividing another one with different rate (e.g. it means sending binary '0').

Method proposed by Servetto et al. [19] which introduces intentional losses in numbered stream of packets may be also utilized. This solution is implemented as skipping one sequence number at the sender so no user data is lost. Loss that occurred during fixed time interval is equal to sending one steganogram bit. This method is called *phantom packets*. The same method can be applied to IP fragmentation. While sender generates fragments, it skips one *Fragment Offset* value and inserts the user data into next fragment. If the loss of fragment occurs it means sending binary '1' and if it is not present, binary '0'. To work correctly this method requires modified receiver which can reassembly original IP packet even though not all fragments reached the receiver. We named this modified version of existing method as *phantom fragments*.

For all presented above methods steganographic bandwidth equals PRBR = 1 bit/packet.

**4.2 PMTUD**

The main idea for exchanging hidden data with PMTUD is simple – it involves transmitter to utilize probe messages to carry steganogram and invoke sending intentional fake ICMP messages by receiver. Detailed exchanging hidden information procedure is suitable for both IPv4 and IPv6 and is possible for all scenarios from Fig. 5.

Proposed steganographic method works as follows. SS knows from previous interactions with SR what the correct MTU for their communication path is. When SS wants to send steganogram then it sends a probe message that contains steganogram inserted into packet payload. The size of the packet is set to the maximum MTU allowed for path between SS and SR, thus SS is certain that this packet will reach the receiver.

We propose similar procedure to make the selected packet for steganographic purposes distinguishable from other yet hard to detect as it was presented for IP fragmentation mechanism. If we assume that sender and receiver share secret Steg-Key (*SK*), then for each packet chosen for hidden communication a hash function (*H*) is used to calculate Identifying Sequence (*IS*):

$$IS = H(SK \parallel Identification \parallel CB) \qquad (4\text{-}3)$$

where *Identification* denotes values from that IP header field, *CB* is *Control Bit* and $\parallel$ is bits concatenation function. *Control Bit* is used to inform the receiver whether it should sent more fake ICMP messages or not (*CB=1* send more ICMP, *CB=0* do not send more ICMP). For every IP packet used for hidden communications the resulting *IS* will be different due to the changing values from Identification field. All *IS* bits or only selected ones are distributed across payload field in predefined manner.

After a probe message reaches the receiver, he/she calculates two *IS*s (one for *CB=1*, second for *CB=0*) based on *SK* and value from the IP header and checks if it contains steganogram or user data. When steganogram is detected it is extracted from the packet payload. If *IS* calculation indicates that *CB=1* then receiver intentionally send ICMP message that indicate that the MTU of the path must be decreased (type=3, code=4) and thus sender is obligated to send smaller probe message (which will also contain steganogram). In fake ICMP message source IP address must be spoofed to avoid trivial detection. In the payload of ICMP message IP header of the original packet and 64 bits of original data are present. Receiver must mark ICMP message to allow sender to distinguish real ICMP from fake one. To achieve this we propose to modify the TTL (Time To Live) field of the original IP packet header from the ICMP payload and change the *Total Length* and *Header Checksum* values accordingly. TTL is the only field in IP header (if IP fragmentation is not used) which may be modified during traversing the network thus comparing original packet sent with returned in ICMP message will not result in easy hidden communication detection. There are many possibilities of TTL modifications and in particular they include setting TTL to prearranged value or to even/odd one. Functioning of the described above steganographic method is also illustrated in Fig. 7. In this example, during the PMTUD exchange, about 3 kB of steganogram was sent from SS to SR.

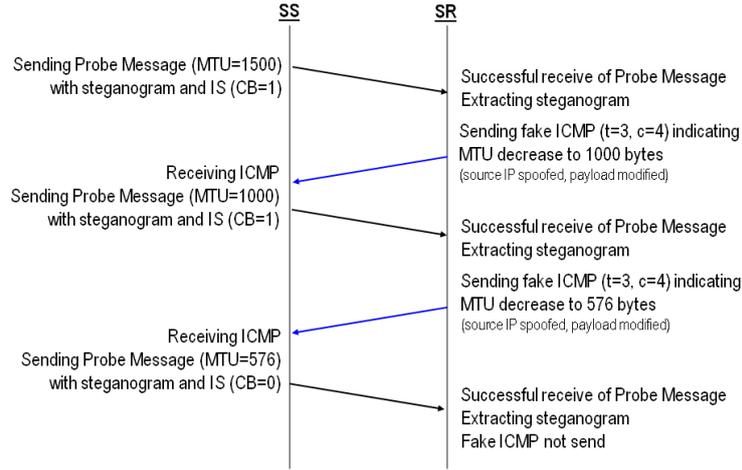

**Fig. 7** PMTUD steganographic method

For proposed method steganographic bandwidth can be expressed with as:

$$RBR_{PMTUD} = \frac{\sum_{1}^{n} P_n}{T} \quad [bits/s] \tag{4-4}$$

where $n$ denotes number of probes sent from sender to receiver, $P_n$ probe payload size and $T$ connection duration.

During PMTUD exchange all probes messages may be used for steganographic purposes but in this case detection may be easier to perform. Because it is assumed that the earlier probes failed to reach the receiver, next ones should carry fragment of the same data. Thus, comparing each probe message sent with the first one issued may be used to detect steganograms. Only in case when the first probe is used to carry steganogram above steganographic method is hard to detect but the steganographic bandwidth is limited.

**4.3 PLPMTUD**

In PLPMTUD probes messages are validated at the transport layer and correct MTU is learned by starting with packets which size is relatively small and when they get through they proceed with progressively larger ones. The isolated loss of a probe packet is treated as an indication of an MTU limit and transport layer protocol is permitted to retransmit any missing data. Thus, steganographic method described for PMTUD is not applicable. Nevertheless, other possibilities for hidden communication may be utilized. One of them is RSTEG (Retransmission Steganography) method which is presented in details in [21] and uses intentional retransmissions to sent steganograms. RSTEG main idea is to not acknowledge a successfully received packet in order to intentionally invoke retransmission. The retransmitted packet carries a steganogram instead of user data in the payload field. RSTEG may be used for IPv4 and IPv6 in all hidden communication scenarios from Fig. 5.

For PLPMTUD using RSTEG works as follows. SS knows from previous interactions with SR what the correct MTU for their communication path is. When the connection starts, SS sends probe message with prearranged MTU. After successfully receiving the packet, the receiver intentionally does not issue an acknowledgment message. In a normal situation, a sender is obligated to retransmit the lost packet when the timeframe within which packet acknowledgement should have been received expires. In the context of RSTEG, a sender replaces original payload with a steganogram instead of sending the same packet again. When the retransmitted packet reaches the receiver, he/she can then extract hidden information.

The detection method is similar to one presented for PMTUD and is based on comparing probes messages payload during MTU learning process.

## 5. Conclusions

In this paper we presented potential steganographic methods that can be used for mechanisms for handling oversized IP packets: IP fragmentation, PMTUD and PLPMTUD. In particular, we propose two new steganographic methods, two extensions of existing ones and we show how IP fragmentation simplifies utilizing steganographic solutions which require transmitter-receiver synchronization.

Proposed methods can be utilized to enable hidden communication for both versions of IP protocol: 4 and 6. They are characterized by different steganographic bandwidth and detection possibilities, thus they can have various impact on network security. Knowledge of these information hiding procedures can be now to develop and implement countermeasures for network traffic monitoring, which may limit the risk of confidential information leakage or other threats caused by covert communication.